\documentclass[final,5p,times]{elsarticle}

\usepackage{hyperref}
\usepackage{blindtext}

\usepackage{eqexpl}
\usepackage{amsmath}
\usepackage{amsfonts}
\usepackage{amssymb}
\usepackage{graphicx}
\usepackage{caption,xcolor}
\usepackage{subfig}
\usepackage{mathtools}
\eqexplSetIntro{where:} 
\eqexplSetDelim{=} 
\newcommand{\ie}{{\em i.e.}}
\newcommand{\eg}{{\em e.g.}}
\newcommand{\apriori}{{\em a priori}}

\begin{document}

\begin{frontmatter}

\title{Beyond the Standard Model in Vector Boson Scattering Signatures}

\author[1]{Michele Gallinaro (ed.)}
\ead[1]{michele.gallinaro@cern.ch}
\address[1]{Laborat\'orio de Instrumenta\c{c}\~{a}o e F\'isica Experimental de Part\'iculas, LIP Lisbon, Av. Prof. Gama Pinto, 2 - 1649-003, Lisboa, Portugal}

\author[2]{Kenneth Long (ed.)}
\ead[2]{kenneth.long@cern.ch}
\address[2]{Experimental Physics Department, CERN, 1 Esplanade des Particules, 1211 Gen\`eve 23, Switzerland}

\author[3]{J\"urgen Reuter (ed.)}
\ead[3]{juergen.reuter@desy.de}
\address[3]{DESY, Theory Group, Notkestr. 85, 22607 Hamburg, Germany}

\author[4]{Richard Ruiz (ed.)}
\ead[4]{richard.ruiz@uclouvain.be}
\address[4]{Centre for Cosmology, Particle Physics and Phenomenology (CP3),\\ Universit\'{e} Catholique de Louvain, 2, Chemin du Cyclotron, B-1348 Louvain-la-Neuve, Belgium}

\author[5]{Dinos Bachas}
\address[5]{Aristotle University of Thessaloniki, Greece}

\author[6]{Liron Barak}
\address[6]{Tel Aviv University, Israel}

\author[3]{Fady Bishara}

\author[7]{Ilaria Brivio}
\address[7]{Institut f\"ur Theoretische Physik, Universit\"at Heidelberg
Philosophenweg 16, 69120 Heidelberg, Germany}

\author[8]{Diogo Buarque Franzosi}
\address[8]{Department of Physics, Chalmers University of Technology,
Fysikg\aa rden, 41296 G\"oteborg, Sweden}

\author[9]{Giacomo Cacciapaglia}
\address[9]{Institut de Physique des 2 Infinis (IP2I), CNRS/IN2P3, UMR5822, 69622 Villeurbanne, France;\\ and Universit\' e de Lyon, Universit\' e Claude Bernard Lyon 1, 69001 Lyon, France}

\author[10]{Farida Fassi} 
\address[10]{Mohammed V University in Rabat, Morocco}

\author[5]{Eirini Kasimi}

\author[11]{Henning Kirschenmann}
\address[11]{Helsinki Institute of Physics, P.O.Box 64, 00014 University of Helsinki, Finland}

\author[5]{Chara Petridou}

\author[12]{Harrison Prosper}
\address[12]{Department of Physics, Florida State University, Tallahassee, FL 32306, USA}

\author[13]{Jorge C.~Rom\~ao}
\address[13]{Departamento de F\'{\i}sica and CFTP, Instituto Superior T\'ecnico\\
Universidade de Lisboa, Av. Rovisco Pais 1, 1049-001 Lisboa, Portugal}

\author[14]{Ignasi Rosell}
\address[14]{Departamento de Matem\'aticas, F\'\i sica y Ciencias Tecnol\' ogicas, Universidad Cardenal Herrera-CEU,\\ CEU Universities, 46115 Alfara del Patriarca, Val\`encia, Spain}

\author[15]{Ennio Salvioni}
\address[15]{Theoretical Physics Department, CERN, 1 Esplanade des Particules, 1211 Gen\`eve 23, Switzerland}

\author[16]{Rui Santos} 
\address[16]{Centro de F\'{\i}sica Te\'{o}rica e Computacional,
Faculdade de Ci\^{e}ncias, Universidade de Lisboa, 
1749-016 Lisboa, Portugal,\\
and
ISEL -  Instituto Superior de Engenharia de Lisboa,Instituto Polit\'ecnico de Lisboa
1959-007 Lisboa, Portugal}

\author[17]{Magdalena Slawinska}
\address[17]{Institute of Nuclear Physics, Polish Academy of Sciences, ul. Radzikowskiego 152, 31-342 Krak{\'o}w, Poland}

\author[1,18]{Giles Chatham Strong}
\address[18]{Universit\`{a} degli Studi di Padova, Via Marzolo 8, 35131 Padova, Italy}

\author[19]{Micha{\l} Szleper}
\address[19]{National Centre for Nuclear Research, Pasteura 7, 02-093 Warszawa, Poland}

\begin{abstract}

The high-energy scattering of massive electroweak bosons, known as vector boson scattering (VBS), is a sensitive probe of new physics.  VBS signatures will be thoroughly and systematically investigated at the LHC with the large
data samples available and those that will be collected in the near future. Searches for  deviations from Standard Model (SM) expectations in VBS facilitate tests of the Electroweak Symmetry Breaking (EWSB) mechanism. Current state-of-the-art tools and theory developments,
together with the latest experimental results, and the studies foreseen for the near future are
summarized. 
A review of the existing Beyond the SM (BSM) models that could be tested with such studies as well as data analysis strategies  to understand the interplay between models and the
effective field theory paradigm for interpreting experimental results are discussed.
This document is a summary of the EU COST network ``VBScan" workshop on the sensitivity of VBS processes for BSM
frameworks that took place December 4-5, 2019 at the LIP facilities in Lisbon, Portugal. In this
manuscript we outline the scope of the workshop, summarize the different contributions from theory and
experiment, and discuss the relevant findings.
\end{abstract}


\end{frontmatter}

\section{Introduction}

This document summarizes the contributions, dis\-cus\-sions, and conclusions from the topical workshop
on ``Beyond the Standard Model (BSM) processes in Vector Boson Scattering (VBS) signatures," of the 
EU COST Action 16108 ``VBScan", which took place on December 4-5, 2019, at the Laborat\'orio de
Instrumenta\c{c}\~{a}o e F\'isica Experimental de Part\'iculas, LIP Lisbon. The main scope of this
workshop was to bring together scientists working on BSM physics within the community of
experimentalists and theorists focused on the VBS signatures at the LHC. The purpose was two-fold: (1) to give an overview over the status of these 
\begin{figure*}[htb]
    \centering
    \includegraphics[width=.95\textwidth]{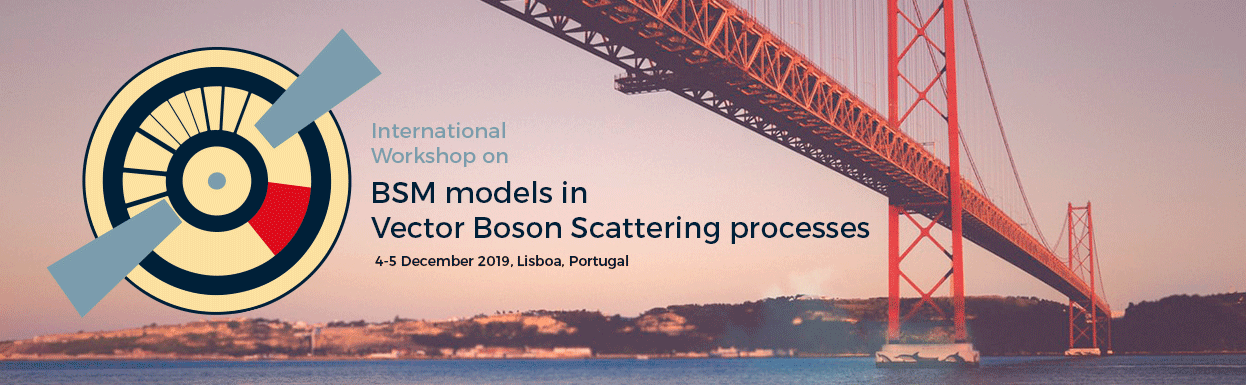}
    \caption{Logo of the VBScan Workshop at LIP Lisbon, December 4-5, 2019.}
    \label{fig:my_label}
\end{figure*}
measurements with the LHC Runs 1 and 2 at $\sqrt{s}=7,$ 8 and 13 TeV, and (2) to use the existing constraints on extensions of the Standard Model (SM) to see whether these signatures can be used to obtain more information on specific BSM models. Phenomenologists working on BSM model-building usually focus on simpler processes, \ie\ with (much) higher cross sections (\eg\ top processes, Drell-Yan, dibosons, Higgs processes), while the efforts of the experimental ATLAS and CMS collaborations for VBS until now only covered measurements of SM cross sections and setting limits on deviations of the SM in terms of dimension-6 and dimension-8 operators in a SM Effective Field Theory (SMEFT or HEFT). The workshop was intended to bring together these two communities, to kick off possible collaborations, to raise the interest of BSM phenomenologists into VBS signatures as well as the interest of experimentalists into specific BSM models that are testable  with VBS data. In particular, discussions were fostered on how to best utilize the Run~2 and upcoming Run~3 data sets for VBS processes. 

This document is structured into four parts: The first part, Sec.~\ref{sec:exp}, 
gives an overview of the existing VBS results in the different channels, including prospects 
for the upcoming LHC Runs. Furthermore, different experimental techniques to minimize systematic 
errors, to use boosted topologies for hadronic channels, etc. are discussed. The next part, Sec.~\ref{sec:theo}, discusses three different classes of BSM setups with increasing definiteness: SM effective field theory (SMEFT), simplified models with specific new heavy states, and explicit models. Section~\ref{sec:synergy} then discusses the synergies between experimental measurements and searches on the one hand and theoretical developments and calculations on the other. Finally, Sec.~\ref{sec:summ} summarizes the discussions and the lessons learned from this workshop. 

\section{Experimental status of BSM searches in VBS}
\label{sec:exp}

 \subsection{Overview of current experimental results}
\label{sec:exp_results}

The experimental aspect of the workshop began with broad summaries of experimental measurements related to VBS.
A summary of the measurements of VBS diboson production in the SM and on searches for BSM physics, and the prospects for 
future improvements and extensions, was presented~\cite{KLong,DBachas}. While the first VBS measurements were made at 8
TeV during the LHC Run~1 ~\cite{Khachatryan:2014sta,Khachatryan:2016vif,Khachatryan:2017jub,Aad:2014zda,Aaboud:2017pds}, 
with the nearly 140 fb$^{-1}$ of data collected by the ATLAS and CMS Collaborations during the LHC Run~2, many VBS processes 
have recently been explored for the first time. However, in many cases the full integrated luminosity of Run~2 has not yet been 
exploited, and new results are expected to arrive in the coming months. Furthermore, the LHC has delivered only a small 
fraction of the total integrated luminosity expected from its current phase and the future high-luminosity upgrade (HL-LHC).

The first VBS measurement exploiting the full Run~2 data set was performed by the
ATLAS Collaboration, studying $ZZjj$ production with the Z boson pair decaying to either four charged leptons (4$\ell$) or two charged leptons and two neutrinos ($2\ell2\nu$)~\cite{ATLAS:2019vrv}.
The large data set is particularly beneficial for this process, which has a very low
cross section, but an extremely clean signature in the four lepton decay channel. A
multivariate analysis was used to fully exploit the data set, which trained separate
boosted decision trees (BDT) for the $4\ell$ and $2\ell2\nu$ signal
categories. In the four lepton channel, a control region was built from events that
do not satisfy either $m_{jj} > 300$\,GeV or $\left|\Delta\eta_{jj}\right| > 2.0$, which are required for signal events. A maximum likelihood fit was performed simultaneously to the $m_{4\ell}$ distribution in this background control region as well as the BDT discriminant score in the $4\ell$ and $2\ell2\nu$ channels to derive the observed signal strength, $\mu = \sigma_{obs}/\sigma_{exp}$ = 1.35 $\pm$ 0.34, where $\sigma_{obs}$ and $\sigma_{exp}$ are the observed and expected cross sections. The observed (expected) significance of this result is quantified with respect to the background-only hypothesis of the SM without VBS ZZ production at 5.3$\sigma$ (3.5$\sigma$), which constitutes the first observation of this process. The sensitivity of the measurement is strongly driven by the four lepton channel.

The CMS Collaboration has also performed a study searching for VBS ZZ production using 35.9 fb$^{-1}$ of data collected in 2016~\cite{Sirunyan:2017fvv}. The results for this analysis were also extracted via a maximum likelihood fit to the distribution of a BDT discriminant score. No control region is explicitly defined, rather, the BDT score is trained and evaluated on a loose selection of events with $m_{jj} > 100$ GeV. Therefore, the low BDT-score regions effectively serve as a control region in the fit. The observed signal strength is reported to be $\mu = 1.39^{+0.86}_{-0.65}$, with an observed (expected) significance of 2.7$\sigma$ (1.6$\sigma$).

The same-sign $W^{\pm}W^{\pm}$ VBS process is widely regarded as the golden channel for experimental measurements, as it is the only VBS process where the electroweak (EW) contribution dominates over the production of dibosons with jets from QCD radiation (QCD production)~\cite{Ballestrero:2018anz}. Due to its striking same-sign lepton signature and low background, it was the first VBS process to be observed at the LHC, 
first by the CMS Collaboration~\cite{Sirunyan:2017ret} and later by the ATLAS Collaboration~\cite{Aaboud:2019nmv}. The two analyses follow a very similar strategy: backgrounds from non-prompt leptons are estimated from control regions in data that consist of events failing lepton identification requirements. The contributions from EW and QCD WZ production in the signal region selection were estimated from MC simulation, but corrected using dedicated three-lepton control regions. The signal strength is extracted via a fit to the $m_{jj}$ spectrum of selected events in the ATLAS analysis, and to a two-dimensional distribution of $m_{jj}$ and $m_{\ell\ell}$ in the CMS analysis. The observed signal strengths are consistent with each other, with the ATLAS (CMS) analysis reporting a significance of 6.5$\sigma$ (5.5$\sigma$) over the null hypothesis. The fiducial cross sections reported are consistent with the SM predictions, though some tension is observed between the ATLAS measurement and the prediction from Sherpa v2.2.2~\cite{Gleisberg:2008ta}. This discrepancy has been understood in terms of the color flow treatment in Sherpa~\cite{sherpaMBITalk}.

The ATLAS and CMS Collaborations have further performed measurements with 36 fb$^{-1}$ of VBS 
WZ~\cite{Aaboud:2019nmv,Sirunyan:2019ksz}, Z$\gamma$~\cite{Sirunyan:2020tlu}, and WZ or ZZ with one boson decaying
leptonically~\cite{Aad:2019xxo,Sirunyan:2019der}, referred to as WV and ZV, where the V = W, Z decays hadronically. The CMS Collaboration uses the final state to probe anomalous $VV$ 
production, whereas the ATLAS Collaboration also performs a search for the SM production. This is accomplished with an 
analysis that considers nine independent signal regions, divided by the decay of the W or Z boson and whether the hadronic decays 
form distinct or merged jets. Independent BDTs are trained for each signal region, which helps the analysis overcome 
the huge background from V+jet processes. The observed production rate with respect to the SM for VBS VVjj production is 
$1.1^{+0.42}_{-0.40}$, corresponding to an observed significance of 2.7$\sigma$ with 2.5$\sigma$ expected in the SM.

The VBS production of a massive vector boson accompanied with a photon was first studied at 8 TeV~\cite{Aaboud:2017pds}. Studies have recently been performed for Z$\gamma$ VBS production at 13 TeV~\cite{Aad:2019wpb,Sirunyan:2020tlu}. The analysis performed by the CMS Collaboration selects events with a leptonically decaying Z boson and a photon associated with two jets, and exploits the kinematic distribution of the mass and rapidity separation of the jet to extract results with a binned maximum likelihood fit. The SM production rate with respect to the SM expectation at LO is measured to be $0.64^{+0.23}_{-0.21}$, with observed (expected) significance of 3.9$\sigma$ ($5.2\sigma$). When combined with the 8 TeV result under the assumption of the SM production rate, the statistical significance of the VBS contribution is $4.7\sigma$ (5.2$\sigma$). The measurement performed by the ATLAS collaboration explores EW Z$\gamma$ production using a maximum likelihood fit to the Zeppenfeld centrality variable~\cite{Rainwater:1996ud} of the Z$\gamma$ system, and using a BDT trained on a larger set of characteristics of the EW Z$\gamma$ process. The BDT-driven analysis is more sensitive, with a measurement compatible with the SM at 4.1$\sigma$ observed and expected significance. It is also highly compatible with the analysis exploiting a single distribution, which provides confidence that the multi-variate approach does not bias the results.

For the majority of these results, only a fraction of the data collected in the LHC Run~2 is analyzed. Improved results exploiting the full data set of nearly 140 fb$^{-1}$ are expected soon, and some preliminary studies have already been released since the time of this workshop~\cite{CMS-PAS-SMP-19-012}. In the longer term, the LHC will be upgraded to the HL-LHC phase that will provide dramatically more data, allowing for a rich characterization of VBS processes as discussed in Sec.~\ref{sec:future}.

\subsection{Overview of BSM searches using VBS events}

The lack of clear signs of BSM physics at the LHC have necessitated looking for hints of new physics that are more subtle or more exotic than assumed by traditional approaches. 
Because VBS is a probe of the SM that is sensitive to modifications of the EW sector, and because it is only now becoming experimentally accessible, it is a natural avenue for experimental searches to expand.
The LHC experiments have built comprehensive experimental programs exploring VBS production in the SM, as discussed in the previous section. However, the properties of many of these rare processes are not yet precisely measured, and as such, a precise comparison of their agreement with theoretical predictions is often not possible. Furthermore, effects from new physics that do not have an appreciable impact on the total production cross section but show some clear signature in other kinematic regions might not be visible in a measurement specifically designed to measure SM VBS production. In such cases, analyses designed to focus on new physics searches are complementary to SM measurements. 

Searches for new physics in VBS channels can broadly be divided into those which look for explicit (but possibly simplified) models of new physics, and generalized searches, usually parameterized in the language of EFT~\cite{DBachas}. The impact of new physics from non-zero dimension-8 operators has been studied by the CMS Collaboration at 13 TeV in the W$^{\pm}$W$^{\pm}$\cite{Sirunyan:2017ret}, ZZ~\cite{Sirunyan:2017fvv}, Z$\gamma$~\cite{Sirunyan:2020tlu}, WZ~\cite{Sirunyan:2019ksz}, and WV/ZV~\cite{Sirunyan:2017fvv} channels, and previously by the ATLAS and CMS Collaborations at 8 TeV. In all cases, events are selected to enhance VBS VV production, and a distribution of events sensitive to the modification of the energy of the scattering, such as the mass of the VV system, is used to place constraints on the operator couplings. While VBS VV production with semi-leptonic decays is a challenging experimental channel, its high branching fraction provides the strongest handle on dimension-8 EFT operators. Exclusive VV production, discussed in Sec.~\ref{sec:exclusive}, gives even stronger results for operators sensitive to the WW$\gamma\gamma$ interaction. A full and up-to-date summary of results is maintained at Ref.~\cite{acSummary}. The following sections discuss the validity and interpretation of these constraints from a theoretical perspective.

In general, modifications of the SM are unlikely to be confined to VBS processes. The EFT operators studied in VBS analyses are also relevant for non-VBS VV production, VVV production, and production of the Higgs boson. New resonances in the EW sector would also likely couple to the vector bosons and Higgs boson such that many other production mechanism would be impacted. Therefore, searches for new physics in diboson and Higgs events have strong implications for new physics searches in VBS channels. The experimental program searching for such processes is exhaustive. An overview of relevant results for diboson resonances, Higgs production via gluon-gluon and vector boson fusion, and double Higgs production is given in Ref.~\cite{DBachas}.

Of the many possible models predicting modifications to the EW sector, searches for additional Higgs bosons are of considerable theoretical and experimental interest~\cite{LBarak}. Depending on the mass and couplings of the new scalar particle, VBS may not be a practical avenue for its discovery. The CMS and ATLAS collaborations have performed many searches using diboson~\cite{Sirunyan:2019gkh,Sirunyan:2019bez,Aaboud:2019gxl,Aad:2019fbh} and leptonic decays of the hypothesized H$^{\pm}$~\cite{Aaboud:2018gjj,Sirunyan:2019hkq}, resulting in strong constraints on its possible mass and couplings. If the H$^{\pm}$ is fermiphobic, VBS would be a principle production mechanism. A well-studied model in which a charged Higgs sector that preserves custodial symmetry is introduced is the Georgi--Machacek (GM) model~\cite{Georgi:1985nv}. The ATLAS and CMS collaborations have performed searches using VBS events for the GM H$^{\pm\pm}$ in the W$^{\pm}$W$^{\pm}$\cite{Sirunyan:2017ret} channel, and for H$^{\pm}$ in the WZ~\cite{Aaboud:2018ohp,Sirunyan:2019ksz} and WV/ZV~\cite{Sirunyan:2017fvv} channels. The H$^{\pm\pm}$ and H$^{\pm}$ have the same mass in the GM model, but the results have not yet been combined across channels or across experiments. However, a small and broad fluctuation seen in the ATLAS WZ VBS analysis is not present in the CMS results~\cite{Aaboud:2018ohp,Sirunyan:2019ksz,Sirunyan:2017fvv}.

\subsection{Exclusive VV production and proton tagging}
\label{sec:exclusive}

Recently, there has been a renewed interest in studies of central exclusive production (CEP) processes in high-energy proton-proton collisions.
A summary was presented~\citep{MGallinaro}, including the experimental challenges, current status and future prospects.
In CEP processes in proton-proton collisions, the exchange is mediated through photon-photon fusion and particle production with masses at the electroweak scale can be studied.
CEP provides a unique method to access a variety of physics topics, such as new physics via anomalous production of W and Z boson pairs, high transverse momentum 
($p_T$) jet production, and possibly the production of new resonances. 
These studies can be carried out in particularly clean experimental conditions thanks to the absence of proton remnants.

Studies of exclusive production can be performed at the CMS experiment by tagging the leading proton from the hard interaction. To this end, the Precision Proton Spectrometer (PPS)~\cite{Albrow:2014lrm} provides an increased sensitivity to selecting exclusive processes. 
The PPS is a detector system to add tracking and timing information at approximately 210~m 
from the interaction point around the CMS detector. 
It is designed to operate at high luminosity with up to 50 interactions per 25~ns bunch crossing to perform measurements of, 
\eg\, the quartic gauge couplings and search for rare exclusive processes. Since 2016, PPS has been taking data in normal high-luminosity proton-proton LHC collisions, and it collected approximately 100~fb$^{-1}$ of data.

CEP of an object X may occur in the process $pp\rightarrow p + X + p$, where "+" indicates the ``rapidity gaps" adjacent to the state $X$. 
Rapidity gaps are regions without primary particle production.
In the high mass region with both protons detected, among some of the most relevant final states are $X = e^+e^-,\mu^+\mu^-,\tau^+\tau^-$ and $W^+W^-$. 
In CEP processes, the mass of the state $X$ can be reconstructed from the fractional momentum loss $\xi_1$ and $\xi_2$ of the scattered protons
by using the expression $M_X=\sqrt{\xi_1\cdot \xi_2\cdot s}$.
The $M_X$ reach at the LHC is significantly larger than at previous colliders because of the larger $\sqrt{s}$.
The scattered protons can be observed mainly thanks to their momentum loss, due to the horizontal deviation from the beam trajectory. 
The acceptance in $\xi$ depends on the distance from the interaction point and on how close to the beam the proton detectors can be moved.
For the first time, proton-proton collisions at the LHC provide the conditions to study particle production with masses at the electroweak scale through photon-photon fusion.
At $\sqrt{s}=13$~TeV and in normal high-luminosity conditions, values of $M_X$ above 300~GeV can be probed.
CEP processes at these masses have small cross sections, typically of the order of a few fb, and thus can be studied in normal high-luminosity fills.

The exclusive two-photon production of pairs of photons, $W$ bosons, and $Z$ bosons, provides a novel and unique testing ground for the electroweak gauge boson sector. The detection of 
$\gamma\gamma\rightarrow W^+W^-$ events allows one to measure the quartic gauge coupling $WW\gamma\gamma$ with high precision. 
One can study the distributions and measure the production rates of these interactions, and verify whether they are compatible with the SM. An improvement in sensitivity of the order of $10^{-3}-10^{-4}$ is expected with respect to earlier measurements~\cite{Chatrchyan:2013akv,Sirunyan:2017fvv,Aaboud:2017tcq}.
As a first step, the exclusive dilepton process $pp \rightarrow p\ell^+\ell^-p^{(*)}$ ($\ell=e,\mu$) 
has been observed for the first time at the LHC in pp collisions at $\sqrt{s}=13$~TeV~\cite{Cms:2018het}.

At large $\sqrt{s_{\gamma\gamma}}$, the two-photon process $\gamma\gamma\rightarrow W^+W^-$ provides a window to BSM physics, since it is sensitive to triple and quartic gauge boson couplings. In pp collisions at $\sqrt{s}=8$~TeV, CMS has observed 13 candidate events in a final state with $e^\pm\mu^\mp$, large missing transverse energy, and no additional tracks, but without detecting the protons~\cite{Khachatryan:2016mud}. 
The observed yields and the kinematic distributions are compatible with the SM prediction for exclusive and quasi-exclusive $\gamma\gamma\rightarrow W^+W^-$ production.
The results are used to derive upper limits on the anomalous quartic gauge coupling (aQGC) parameters.
With an integrated luminosity of 100~fb$^{-1}$, the PPS is expected to improve the limits by at least two orders of magnitude, or perhaps observe a deviation from the SM production.

Among other interesting topics, the PPS can also probe the presence of composite Higgs and anomalous gauge-Higgs couplings, 
search for excited leptons, technicolor, extra-dimensions, axions, heavy exotic states, dark matter candidates, 
and explore more BSM processes~\cite{Delgado:2014jda,Inan:2010af,Fichet:2013ola,Fuks:2019clu}.

\subsection{Machine learning in measurements and searches}

Because the amount of data collected at the LHC will multiply at a much slower rate in the coming years, innovative experimental techniques are crucial for the future success of the field. Outside of particle physics, the attention and value placed on data analysis has increased dramatically in the past years. As such, there are many tools developed independently of the field that may potentially be valuable assets to particle physics. 

A broad class of data analysis tools referred to as Machine Learning (ML) leverage computational algorithms to identify and exploit features in a data set. An overview of the broad scope of ML was presented~\cite{GStrong}, including examples of its use, and benefits in physics analyses. For example: in signal-versus-background discrimination in particle physics analyses, \emph{Supervised learning} can be used. In such a case, MC simulations are used to define expected behaviors of  signal and background processes, and the ML algorithm serves to build a function (\ie, a model) predicting whether the attributes of an event, commonly referred to as \emph{features}, are most likely associated to signal- or background-like processes. Boosted decision trees (BDT) and neural networks (NN) are two widely studied ML models that are increasingly adopted for use in analysis and reconstruction in particle physics. They serve as complex and flexible functions that are ``trained," that is, statistically fitted, to describe the data based on features in an automated way. Training is built around the minimization of a \emph{loss function} which quantifies the ability of the model to describe the training data. This approach provides the opportunity for a more extensive and less manual optimization procedure than traditional selection-based approaches, where the ``feature engineering'' is performed manually by a physicist.

ML approaches are already used widely in LHC analyses, including in VBS measurements and searches. It is also expected that their use will increase at the HL-LHC, where the higher pileup environment will make reconstruction more challenging. In particular, the complexity of combinatorial algorithms used to build tracks from ``hits'' scale dramatically as the number of hits increases, whereas ML algorithms can provide nearly constant run-time. Likewise, searches for very rare phenomena, such as di-Higgs production, will require maximal separation of a very small signal from huge backgrounds. The prospects for this analysis have recently been studied by the CMS Collaboration using a deep NN which provides enhanced sensitivity over traditional selection-based approaches~\cite{CMS-PAS-FTR-18-019}.

\subsection{Future prospects for experimental VBS studies}
\label{sec:future}
The outlook for future measurements of and searches for VBS processes is promising, particularly on more immediate timescales. In 2021, Run~3 of the LHC program will start and is estimated to deliver $\mathcal{L}\approx 300$ fb$^{-1}$ of data to each of the ATLAS and CMS experiments.
Following this period are Run~4 and subsequent runs of the LHC program,
that is to say the HL-LHC phase.
Prospects for the HL-LHC, which is slated to deliver 
around $\mathcal{L}=4.5-5$ ab$^{-1}$ of data,
are extensively documented in community reports~\cite{Dainese:2703572,Cerri:2018ypt,CidVidal:2018eel}.

Opportunities and possibilities for VBS beyond the HL-LHC
are also actively being discussed in community-wide exercises, such as the ``European Strategy Update,'' and the analogous ``Snowmass''  process in North America.
Present benchmarks consider a $\sqrt{s}=27$ TeV upgrade of the LHC, a prospect known as the High Energy Large Hadron Collider (HE-LHC)~\cite{Cerri:2018ypt,CidVidal:2018eel,Benedikt:2018csr,Abada:2019ono}, as well as a future $e^+e^-$ collider (FCC-ee), and a $\sqrt{s}=100$ TeV circular $pp$ collider (FCC-hh)~\cite{Golling:2016gvc,Mangano:2016jyj}.
While many sensitivity estimations for SM measurements and BSM discovery prospects are reported in these documents, the situation remains dynamic and evolving as future collider outlines mature and become more refined.

\section{Theoretical motivation and precise predictions for BSM physics in VBS}
\label{sec:theo}

This section gives an overview of the theoretical contributions to the BSM Lisbon workshop. They fall into four 
different categories, either presenting topics in one of the three different parameterizations of BSM physics in VBS, or 
discussing the progress in the theoretical description of the SM signal processes. The latter topics have been a separate effort 
within the VBScan COST action and,  via dedicated workshops, led to a publication regarding the precision description of 
the like-sign VBS process, $pp \to jj e^+ \nu_e \mu^+ \nu_\mu$~\cite{Ballestrero:2018anz}. Nevertheless, a precise understanding
of the underlying SM processes is indispensable for a significant discovery of new physics in any channel, and VBS is no exception.
So, these topics have been included in the workshop. They are summarized in subsection~\ref{sec:sm_vbs_proc}.

There are three different layers of definiteness for the parameterization of BSM 
physics in VBS (and generally in other LHC processes):
(1) the semi-model independent description in terms of an SM effective 
field theory (SMEFT or HEFT), which is covered in subsection \ref{sec:smeft}; (2) simplified 
models, which cover the dominant effects of a general BSM physics setup for VBS, 
are discussed in subsection~\ref{sec:simplified}; and (3) UV-complete models, 
summarized in subsection~\ref{sec:uv_complete}.

\subsection{Effective field theories}
\label{sec:smeft}

Discussions at the meeting on the topic of effective field theories (EFTs) began with a general introduction into the par\-a\-digm~\cite{IBrivio}. While it is possible to formulate different EFTs with the same field content, they are nevertheless viewed as the most general, low-energy extensions of the SM when one includes the tower of all higher-dimensional operators built from SM fields respecting the  symmetries of the SM. Under this formulation, gauge symmetries are then valid up to all 
orders in the expansion, while global symmetries like lepton number conservation are only 
accidental symmetries at the lowest order(s). Indeed, there are two different EFTs depending 
on the assumptions made about the $m\approx125$ GeV scalar state discovered in 2012~\cite{Aad:2012tfa,Chatrchyan:2012ufa}. Just 
describing the longitudinal modes of $W^\pm$ and $Z$ as a non-linear $\sigma$-model for 
Goldstone bosons and adding a single scalar particle leads to the so-called Higgs EFT 
(HEFT)~\cite{Alonso:2012px,Brivio:2013pma,Buchalla:2013rka,Gavela:2014vra,Brivio:2016fzo}, leaving the Higgs and the Goldstone bosons theoretically unrelated. Assuming that 
these four states together make up an S$U(2)_L\otimes$U$(1)_Y$ EW doublet lin\-ear\-izes the 
Goldstone boson interactions and leads to an EFT called SMEFT. The non-linear HEFT contains 
SMEFT as a special case and hence is more general. HEFT matches the case of composite Higgs 
and some little Higgs models, and can account both for possible non-linear effects in the 
Higgs sector as well as mixings of the Higgs field with a singlet scalar. Before the 
discovery of the $m\approx125$ GeV state, an EFT with only the non-linear sigma model 
for the Goldstone bosons and other SM interactions, called the electroweak chiral 
Lagrangian~\cite{Appelquist:1980vg,Longhitano:1980iz,Longhitano:1980tm,Appelquist:1993ka,Feruglio:1992wf}, was widely used. Due to the non-linear structure of the Goldstone-boson 
interactions, dimension-6 operators in HEFT contain terms that appear, \eg, in dimension-8 
operators of SMEFT. There was a separate presentation on the connection between the non-linear 
EFT and BSM models ~\cite{IRosell}. The phenomenological impact of the inclusion of a light Higgs boson into the non-linear setup has been studied in~\cite{Pich:2012dv,Brivio:2013pma,Buchalla:2015qju,Corbett:2015mqf,Brivio:2016fzo}, with constraints from 
electroweak precision observables (EWPO) derived in~\cite{Pich:2013fea} and other low-energy 
experiments in~\cite{Pich:2015kwa}. An important step is the matching of such an EFT setup to 
high-scale models as it was done \eg\ in~\cite{Pich:2016lew,Krause:2018cwe,Pich:2020xzo}. An exemplary 
study on the interplay between resonances and the non-linear EFT for 1 TeV lepton colliders 
where no issues with unitarity arise was made in~\cite{Beyer:2006hx}.

All such EFT expansions assume that there are no other light degrees of freedom with masses similar to those of the SM particles (there are certain, well-defined exceptions that are parameterized by very specific phenomena like \eg\ invisible Higgs decays). The expansion parameter of the EFT series is the ratio of typical particle momenta over a general high-energy scale $\Lambda$, where the operator (Wilson) coefficients are numbers assumed to be of order unity divided by the corresponding powers of this scale $\Lambda$. This is the bottom-up approach of EFTs.

There is also the top-down approach which means starting from a UV-complete theory with new 
heavy degrees of freedom (usually in the TeV or multi-TeV range). Integrating out these 
resonances leads to a specific EFT where the coefficients of the operators can be predicted or 
calculated from the UV-complete theory. Following the decoupling theorem~\cite{Appelquist:1974tg}, the renormalization-group flow~\cite{Wilson:1973jj} of 
the higher-dimensional operators (for $d > 4$) guarantees that the SMEFT shares the same IR 
physics as the SM~\cite{Weinberg:1978kz}. EFTs are powerful tools as they are consistent quantum field theories (QFTs) that allow the systematic  calculation of radiative corrections, 
and even work if the UV-complete theory is non-perturbative (in the strong coupling sense). 

One of the main reasons for the revival of EFTs in the past years is the lack of discoveries 
of new particles at the LHC and the entrance into the high-luminosity phase of the LHC 
effectively turning the machine into an intensity frontier instrument. 

In general, dimension-6, SMEFT operators are the leading deformations of the SM (neglecting 
baryon and lepton number-violating operators of odd dimensions like the Weinberg operator at 
dimension-5). However, this power-counting depends on the UV completion. Regarding UV models with new di- or multi-boson resonances, dimension-6 operators in multi-boson final states usually originate from loop corrections of these new heavy degrees of 
freedom to SM observables, while dimension-8 operators originate from tree-level exchange of these heavy degrees of freedom. This setup sometimes leads to the fact that dimension-8 
operators are even the leading BSM effect in processes like VBS. Dimension-6 operators should 
nevertheless not be neglected~\cite{Gomez-Ambrosio:2018pnl}. However, the standard paradigm is 
that for a process like VBS one assumes that any such dimension-6 contributions have been 
measured elsewhere more precisely (\eg\ diboson processes). Hence, one defines the SM augmented 
by (certain) dimension-6 operators as the signal model, and then looks for deviations in terms of 
Wilson coefficients of dimension-8 operators. 

The basis for these operators is arbitrary and physics does not depend on the choice of basis, 
but  a complete basis in a fixed order of the expansion is necessary. Of course, some results 
are much simpler in a certain basis, or effects are easier to calculate. There are also 
equivalence relations among operator bases known as ``re-parameterization invariances,'' some 
of which are realized by integration by parts, or equations of motions, or identities of the 
underlying symmetry algebras. The most widely adopted basis for dimension-6 operators is the 
so-called Warsaw basis~\cite{Grzadkowski:2010es}. For dimension-8, at the time of the workshop no complete basis had been known, but the most important operators for VBS had been classified~\cite{Eboli:2006wa,Degrande:2013kka}, and there were well-defined procedures on how to get a complete basis~\cite{Lehman:2015via,Henning:2015alf,Kobach:2016ami,Gripaios:2018zrz}. In the meantime, a complete list has been provided in~\cite{Li:2020gnx,Murphy:2020rsh}. 
The completely general basis comprises of 2,499 operators at dimension-6, and 36,971 at 
dimension-8, not considering baryon number violation. There are many tools for the use of EFTs like SMEFT for phenomenological collider 
physics (like VBS processes), about which the authors of~\cite{Durieux:2019lnv} give a good 
overview, with many references of tools therein. Dedicated SMEFT model implementations are 
available in~\cite{Brivio:2017btx,AguilarSaavedra:2018nen}.

Though IR divergences are the same in EFTs and their generating UV-complete theories, UV divergences might 
appear differently and could lead to different regularization and renormalization schemes between EFT and 
UV-complete models~\cite{Passarino:2019yjx,deBlas:2017xtg}. There is a plethora of different aspects about the applicability and validity of EFTs in general, and SMEFT in particular~\cite{MSzleper,JReuter}. 

First of all, there is the 
question whether to consider the linear case (\ie\ the insertion in interference terms with the pure SM 
amplitude, leading to terms that are linear in the EFT power counting), or to consider quadratic terms as 
well. There have been studies, \eg~\cite{AguilarSaavedra:2018nen}, showing that linear expansions could
lead to negative fiducial cross sections, marking a breakdown of the EFT, or a region where at least 
higher-dimensional operators have to be considered as well. 

Then, experimentally, the biggest conundrum is between global fits taking into account all possible deviations by higher-dimensional operators versus variations of only a single or at maximum two operator coefficients. The first approach demands to have $\mathcal{O}(20-30)$ 
parameters under control, which turned out to be an important part of the precision Higgs and electroweak 
physics program during the European Strategy Update of Particle Physics 2019~\cite{Strategy:2019vxc}. The 
second approach is far easier, particularly for channels that are severely statistics limited. One of the 
interesting questions is whether it is possible to learn something about UV physics once a sound, 5$\sigma$ 
discrepancy from the SM has been established. For these reasons, ATLAS and CMS are mostly sensitive to 
rather large values of the combination of Wilson coefficient and scale, $C_i/\Lambda$. This, in turn, either 
means a very low scale, so that new physics is probably directly in the kinematic reach of LHC (which, however, does not 
necessarily mean a discovery, especially if new physics comprises a very broad resonance, cf. next section), or 
that operator  coefficients are larger than allowed even for strongly coupled models. Both would push us outside 
EFT domains of validity. On the other hand, for many scenarios, EFT validity restrictions force  operator coefficients to be so small that they are experimentally not detectable, not even potentially with the HL-LHC. In 
Ref.~\cite{Kalinowski:2018oxd,Kozow:2019txg,Chaudhary:2019aim}, ``EFT triangles'' are constructed in a 
plane with the Wilson coefficient(s) and the EFT scale $\Lambda$ on the $x$- and $y$-axis, respectively. 
The upper region is forbidden by unitarity (for every scale there is a maximally allowed Wilson coefficient). The left region is undetectable because the Wilson coefficients are too small, and in the right region too large Wilson coefficients invalidate an EFT expansion. Only a triangle is left where both the EFT is valid, and a signal is large enough to be detectable by the LHC experiments. For some parameter space regions of some models, these triangles vanish completely; in such setups, an EFT description is not useful at all. 

The next issue is, that in order to look for deviations or set exclusion limits, one needs to define signal models
for, \eg\ SMEFT, that are physically meaningful. The relevant scale in VBS events is given 
by the invariant mass of the diboson system, which is only experimentally accessible for the 
fully leptonic $VV\to ZZ$ process and the semi-leptonic $WZ$ mode in the boosted regime 
(there are plans to look into fully hadronic decays with boosted techniques). For the signal 
models, the Monte Carlo (MC) truth information is available, so this scale is accessible. There 
are several procedures to treat events that would exceed scales allowed by unitarity 
constraints for $2\to 2$, $VV\to VV$ scattering amplitudes, for $V=W,Z,H$: (1) Do nothing. 
For such signal models, cross sections are not bound by unitarity limits, and so there is no 
quantum field theory that could result in such a bin-wise yield of signal events. Limits 
taken from such na\"ive signal models clearly give unrealistically optimistic bounds. (2) 
Generating signal events with ``event clipping,'' which entails dropping individual events  
whenever the scale of an event's diboson system exceeds the unitarity limit from the 
corresponding partial scattering wave. This corresponds to a vertex insertion with a 
momentum step function. It ensures consistency with unitarity of $S$ matrices, but being 
non-continuous cannot be derived from a genuine quantum field theory. (3) Use a form factor 
regularization~\cite{Hagiwara:1986vm,Hagiwara:1993ck}. Here, the (squared) amplitude reaches 
a saturation point at the unitarity bound and is then damped by a power law at high energies. This method has two free parameters: a cut-off scale (\apriori\ unrelated to the 
EFT expansion scale) and the exponent $n$ of the multipole ($n$-pole) form factor. Though 
this seems {\em ad hoc}, behavior like this can be observed in strongly coupled systems with broad 
resonances like in pion and kaon physics. (4) Use the so-called $K$- or $T$-matrix 
unitarization (cf.~\cite{Alboteanu:2008my} for the EW chiral Lagrangian, 
and~\cite{Kilian:2014zja} for SMEFT). This is a projection back onto the Argand circle for 
elastic unitary scattering amplitudes. It does not have any free parameters, and can be 
generalized to intrinsically complex amplitudes. This unitarization leads to a saturation of 
the unitarity bound, and hence gives, bin-per-bin, the largest number of signal events allowed 
in any sensible UV-complete quantum field theory. On the other hand, it is the 
maximally optimistic physical signal model. For transversely polarized gauge bosons, this 
unitarization is also possible, but technically more involved because one has to project to 
the different helicity eigenstates~\cite{Brass:2018hfw}. There are also other unitarization  
models, cf. \eg~\cite{Delgado:2015kxa}, that try to relate unitarity constraints to the 
existence of new resonances. This assumes that the strong dynamics behaves in the same or in a 
very similar way to quantum chromodynamics. Decorrelating new resonances from the 
unitarization of higher-dimensional operator insertions leads to a setup of simplified 
models that are discussed next. 

Lastly, besides the constraints on the size of Wilson coefficients from the point of view of an asymptotic expansion and the unitarity of scattering amplitudes, there are also constraints from the possible UV embedding of EFTs that lead to the so-called ``positivity constraints" on linear combinations of Wilson coefficients, cf. \eg~\cite{Zhang:2018shp,Bi:2019phv}.

\subsection{Simplified models}
\label{sec:simplified}

As the next, more specialized parameterization of new physics beyond the rather generic EFT parameterization, one can set up simplified models for VBS processes. These consist of the SM coupled 
to additional resonances to the diboson system. The philosophy behind these simplified 
models is that any enhancement in the high-energy tails of VBS (in this case) observables 
can be understood as the onset of a new resonance that is just at or near the 
kinematic reach of the LHC. Typical examples of full models in which such resonances can 
appear are extended scalar sectors, like two- (or multi-) Higgs doublet models (2HDM), Higgs 
singlet extensions, Higgs triplet extensions (\eg\ the GM model), Little Higgs 
models, supersymmetric models, twin Higgs models, Randall-Sundrum and other 
extra-dimensional models. Several examples of these will be discussed in the next section. 

An educational example of how to derive an (SM)EFT from such a model can be found for Little 
Higgs models in~\cite{Kilian:2003xt}. Decomposing the (unbroken) quantum numbers of the SM 
in the high-energy limit, the electroweak symmetry and the approximate custodial symmetry, 
for the diboson system, one finds that spin-0, spin-1, and spin-2 resonances could couple to 
the diboson system. These resonances can be either singlets, triplets or quintuplets of weak 
isospin (custodial SU$(2)_c$). A singlet scalar resembles that of the resonance found in the 
Higgs singlet extension (and other models), the triplet of that found in the GM 
model, and the quintuplet of that in the Littlest Higgs model. Spin-1 isovector resonances 
are the $\rho$ resonances of composite Higgs models, while a Kaluza-Klein graviton is the 
prime example for an isosinglet spin-2 resonance. The case  for spin-1 resonances is 
intricate as they can mix (after EW symmetry breaking) with the $W$ and $Z$. Usually, one 
makes the assumption that the coupling of these resonances to the SM fermions are almost 
negligible in order to avoid bounds from low-energy experiments and LHC Drell-Yan searches. 
For extended Higgs sectors, Randall-Sundrum, Little Higgs {\it etc.}, this is generally a valid 
assumption.

Such simplified models for VBS have been studied in~\cite{Kilian:2015opv} for resonances 
coupled to the Goldstone boson system (\ie\ the longitudinal modes), and to resonances 
coupled via EW gauge interactions in~\cite{Brass:2018hfw}. This framework allows one to 
treat both weakly and strongly coupled new physics. Integrating out resonances in simplified 
frameworks gives back Wilson coefficients for the scalar, mixed, and transverse (S,M,T) 
dimension-8 operators. Each resonance adds two free parameters, either its mass and 
its coupling to the diboson system, or its mass and width. In general, due to a simple 
counting of degrees of freedoms, the higher the spin, the larger the effect of the 
resonance, \eg\ a larger resonance peak cross section. Generating signal model events with 
tensor resonances is rather intricate, as they contain many redundant (gauge) degrees of 
freedom: a symmetric real tensor has 10 components, of which only the five spin modes are 
physical. The tensor contains a vector field and two scalars as ghosts that ensure 
transversality and tracelessness of the symmetric tensor. Contributions from  unphysical 
degrees of freedom cancel out, but can lead to numerical inefficiencies/instabilities in a 
signal MC simulation. Ref.~\cite{Brass:2018hfw} shows how one can get a very stable 
simulation by explicitly subtracting these degrees of freedom during MC event generation. An 
explicit implementation is available in the~\texttt{WHIZARD}~\cite{Kilian:2007gr} MC tool. 
Other VBS/VBF tools that implement unitarization are 
\texttt{Phantom}~\cite{Ballestrero:2011pe} and \texttt{VBFNLO}~\cite{Perez:2018kav}. As a 
general rule, resonances that couple via gauge interactions to transverse  degrees of 
freedom of EW bosons are much narrower than resonances that have couplings to the Goldstone 
boson sector, which are numerically more substantial for TeV-scale masses. For broad 
resonances, the onset of the resonance just outside the kinematic reach of the LHC very much 
resembles those contributions from dimension-8 operators, amounting mostly to a larger 
normalization of the highest-energy bins. 

Besides VBS, dimension-8 operators or new resonances coupling to two EW bosons that lead to deviations in 
the quartic gauge couplings of the SM not only play a role in VBS but also in EW triboson production, for 
which there is now evidence at the LHC experiments~\cite{Sirunyan:2017lvq,Aaboud:2017lxm,Aad:2019dxu}. 
However, extracting constraints on new physics from unitarity of the scattering amplitudes is much more 
intricate than for VBS. This is work in progress~\cite{Bahl:2020}. Not only is the experimental signal 
different for triboson channels, but from the theory side, quartic interactions are tested in VBS with two 
initial space-like bosons and two on-shell time-like ones whereas in triboson production one has three 
on-shell time-like ones and a very far off-shell, time-like initial one. This offers different kinematic 
information. 

We now turn to the discussion of UV-complete models (or almost complete models) relevant for 
EW multi-bosons physics and VBS.

\subsection{UV complete or partially complete models}
\label{sec:uv_complete}

One of the biggest advantages of the EFT ansatz or simplified model framework is that they are very general 
and cover close to every deformation of the SM that is consistent with the principles of quantum field 
theory. However, this asset is on the other hand also its biggest drawback: it is in general not possible to 
fold in constraints from other sectors or other measurements, because one either has to use the full EFT with 
all operators or embed the simplified model in a full theory. Furthermore, once a 5$\sigma$ discrepancy from 
the SM is established, it is very hard to reconstruct a complete model from a parameterization of the 
deviation in terms of an EFT.

Ultimately one has to see which new physics model fits the data best. The theory part of the workshop started with a general overview of multi-boson measurements (diboson, VBF, and VBS) relevant for BSM physics at the LHC~\cite{FBishara}. A first look at the diboson measurements at LHC shows a success story for EFT: existing limits from LEP are now superseded not because the measurements at LHC are more precise but because the energy rise of deviations from dimension-6 operators gives the LHC quite a lever-arm. Bounds on anomalous triple gauge couplings have been pushed from percent to per-mille level~\cite{Butter:2016cvz,Azatov:2017kzw,Grojean:2018dqj}. These 
diboson channels, now with LHC data, can be also enlarged by the $WH$ and $ZH$ 
channels~\cite{Franceschini:2017xkh,Bishara:2020vix}. Not yet accessible with current data,  one of the most 
interesting future prospects is the sensitivity to di-Higgs production (especially in the VBF setup), which 
can add to searches for new physics via concrete models or via the ``BSM dictionary''~\cite{Bishara:2016kjn}.
Measurements of angular correlations of leptons can be used to constrain completely transverse 
operators~\cite{Panico:2017frx}. All these constraints can be translated via the EFT dictionary relatively 
straightforwardly into bounds on the parameter space of any BSM models with deviations in the EFT sector 
(\ie\ new particles that give contributions to Wilson coefficients of dimension-6 operators at the one-loop 
level). Another interesting example is the occurrence of new axion-like particles (ALPs) that are either 
connected to the CP problem of QCD, dark matter, or in general any kind of spontaneously broken $U(1)$ 
symmetry. Again, these could be detected via contact interactions in the high-energy tails of diboson 
distributions~\cite{Gavela:2019cmq}.

One of the still relatively new paradigms are BSM models of ``neutral naturalness.'' The naturalness issue is 
tied to the fact that any heavy particle coupled to the scalar sector of the SM (even faintly through higher 
loop corrections) tends to drive the Higgs mass to the mass scale of these new degrees of freedom, as long as 
there is no symmetry protecting it. General paradigms for such symmetries include either global inner 
symmetries, as in compositeness models (cf. below), or space-time symmetries, like supersymmetry  in the 
MSSM or NMSSM, and mostly predict colored particles that share certain properties with the top quark to 
cancel the top loop contributions inside the Higgs potential (``top partners''). Under the assumption that 
this cancellation is between terms of similar size, which is the case for these new symmetries, 
colored top partners should not be too much heavier than the top quark. This expectation, however, 
contradicts the null results from LHC searches. This started a model building activity towards models with color-less top partners ("neutral naturalness").

Though there are interesting links to cosmology, collider searches are the most promising avenue for these models~\cite{ESalvioni}. The color factor of the top quark can be accidentally cancelled by a 
degree of freedom in a fundamental representation of any SU$(3)$ symmetry. This is realized in the so-called Twin Higgs 
models~\cite{Chacko:2005pe} where the top partners are complete SM singlets. So in general there is a more or less 
complete copy of the SM (a ``mirror world'') that talks to the SM only via the scalar/Higgs sector. If the twin sector contains
new cosmologically relevant degrees of freedom, like the mirror photon and twin neutrinos, there are bounds from 
large-scale structure~\cite{Chacko:2016hvu}. A light pseudo-Nambu-Goldstone Higgs like in composite models (cf. below) 
serves as a portal to this twin sector. This induces specific modifications of the Higgs couplings and leads to Higgs 
decays into long-lived particles~\cite{Craig:2015pha}. Other portal models with weakly coupled UV completions lead to 
new relatively light and narrow scalar states (radial modes of coset spaces, cf. \eg~\cite{Chacko:2017xpd}). 

At the LHC, the best search is via the radial scalar mode $(\sigma)$ decaying into two EW vectors, $\sigma \to 
VV$~\cite{Buttazzo:2015bka}. In contrast, for strongly-coupled UV completions (composite Twin Higgs) the radial 
$\sigma$ mode is rather heavy and broad~\cite{Barbieri:2015lqa,Low:2015nqa}. For VBS processes, the effects of such 
Twin Higgs radial modes are rather similar to the corresponding ones for standard composite Higgs 
models~\cite{Contino:2011np}. Generally, for composite Higgs models, lower bounds on the compositeness scale $f$ from 
Higgs coupling measurements and EWPO push new resonances visible in VBS outside the reach of LHC and into the realm of 
FCC-hh or comparable future hadron colliders. For composite Twin Higgs models (with a moderate fine tuning) still much 
lower scales are possible~\cite{Contino:2017moj}. Gluon fusion is still the preferred mode for production of the radial 
mode, however, VBS can add important information on the couplings (to vector bosons). 

There are also supersymmetric variants of neutral naturalness, including the so-called tripled top 
model~\cite{Cheng:2018gvu}. This variant has two copies of the MSSM top sector combined with additional 
$Z_2$ and $Z_3$ discrete symmetries. These models have spectra with colored stops at several TeV, 
SM-singlet  stops at a few 100 GeV, and EW-doublet supermultiplets at roughly half a TeV (and eventually 
glueballs of hidden/Twin color at a few 10 GeV). The largest LHC production cross sections are for the EW 
doublets, whose decays to visible particles is model-dependent. Choosing a completely different 
model-building implementation of neutral naturalness can also give SM-singlet scalar partners, the 
so-called hyperbolic Higgs~\cite{Cohen:2018mgv}. These scalars can be probed in VBF processes together with 
missing energy which turns out to be more sensitive than monojet and $ttH$ 
searches~\cite{Craig:2014lda,Ruhdorfer:2019utl}. There are also proposals to use loop-induced processes to search for 
effects from such models~\cite{Craig:2013xia,Goncalves:2017iub,Englert:2019eyl}. VBF processes are also important for 
the discovery potential of pNGB dark matter candidates (e.g. in neutral naturalness)~\cite{Ruhdorfer:2019utl}. 

In general, BSM models that predict deviations, modifications, or additional degrees of freedom that couple to the EW 
sector, are eligible to be scrutinized in VBS topologies at the LHC. This covers a plethora of different models like 
extended Higgs sectors, including Higgs singlet extensions, multi-Higgs doublet models, Higgs triplet models, 
supersymmetric versions thereof, or models with pseudoscalar  particles. In addition many models with new neutral or 
singly-charged vector bosons fall into this category as they can mix (before or after EWSB) with the SM vector bosons 
and can be produced in VBF or VBS topologies. Clearly, this is a vast model space that could not be covered in a 
specialized workshop like the COST meeting at LIP Lisbon. There were, however, several dedicated talks, on the theoretical status of Two Higgs doublet models (2HDM) and variants thereof~\cite{RSantos}; on experimental searches for singly charged Higgs bosons in the GM model and  doubly charged Higgs bosons in the Type II Seesaw neutrino mass model~\cite{LBarak}; on composite Higgs models~\cite{GCacciapaglia,DBuarqueFranzosi} and new polarization features in the MC event generator {\texttt{MadGraph5\_aMC@NLO}}; on neutral naturalness~\cite{ESalvioni}; and on modeling recommendations for doubly charged Higgs production from VBF and other LHC production mechanisms calculated at NLO+PS~\cite{RRuiz}.

The main idea of composite Higgs models is to dynamically break EW symmetry by a vacuum condensate misaligned with the 
EW vacuum, creating a hierarchy between the EW scale $v$ and the compositeness scale $f$, $v = f \sin\theta_c$, with $\sin\theta_c$ the misalignment angle. The SM Higgs boson is then much lighter than other composite excitations because it is a pseudo-Nambu-Goldstone boson (pNGB). Composite Higgs models are classified according to their broken global symmetry groups $G/H$, examples that have been discussed at the Lisbon COST workshop can be found in~\cite{Gripaios:2018zrz,Barnard:2013zea,Cacciapaglia:2014uja}. 

In composite Higgs models, there are modifications to the Higgs coupling to SM vector bosons by the misalignment angle 
$\theta_c$, which is constrained to satisfy $\cos\theta_c \gtrsim 0.9$ from EW precision observables (EWPO) as well as 
Higgs coupling measurements. These constraints could be relaxed when including contributions from additional composite 
scalar and vector resonances~\cite{BuarqueFranzosi:2018eaj}. One of the strongest constraints on models of 
compositeness are the operators that generate SM fermion masses, as they very easily induce flavor-changing neutral 
currents (FCNC) or other flavor-specific processes. Such predictions are in contradiction to many measurements from 
flavor- and other low-energy experiments. 

In generic composite Higgs models, the operator generating the Higgs potential is related to those operators generating 
(non flavor-diagonal) quartic fermion contact interactions. The necessity to have a large enough quartic Higgs 
coefficient in order to generate a $m\approx125$ GeV state induces too large FCNC Wilson coefficients at the same time. 
So-called partial compositeness (PC) disentangles the power counting for the Higgs quartic and fermion mass operators, 
and vastly relaxes the bounds. One interesting signature for composite Higgs models is the scattering of Goldstone 
bosons, so VBS {\em sui generis}. This has been studied for low compositeness scales $f \gtrsim 550$ GeV as well as for scales in the (multi-)TeV range for FCC-hh.

Goldstone boson scattering in Compositve Higgs models works very similarly to pion scattering: it is unitarized by 
composite resonances as well as a broad continuum. Both cases show behavior similar to the case of the simplified 
models and the $T$-matrix unitarization discussed in the previous section, and have been studied for composite Higgs 
models in~\cite{BuarqueFranzosi:2017prc}. Assuming a misalignment angle compatible with EWPO, unitarity of amplitudes 
requires scalar resonances below $m_\sigma \leq 4$ TeV and vector resonances below $m_\rho \leq 13$ TeV, or a broad 
continuum that behaves like a $T$-matrix setup.  Such nearly conformal dynamics can exhibit $\sigma$-like $0^+$ 
resonances~\cite{Hasenfratz:2016gut,Elander:2017cle}. Polarization measurements (angular correlations) will help to 
disentangle the quantum numbers of deviations to be found in VBS. The next section will discuss some of the 
technicalities how to do such a signal model description including polarization~\cite{BuarqueFranzosi:2019boy}.

There are other interesting VBS or VBF channels, \eg\ $VV \to hh$, already discussed above, or $VV \to \eta\eta$, where 
$\eta$ is a possibly relatively light (tens of GeV) pseudoscalar or axion-like particle, e.g. resulting from an anomalous U$(1)$ symmetry. These particles appear in Extended Technicolor (ETC) cases~\cite{Arbey:2015exa} or Little Higgs 
models~\cite{Kilian:2004pp,Kilian:2006eh}. They naturally have small couplings to SM fermions, and hence their best 
search channels are VBF or gluon fusion (like in PC~\cite{Cacciapaglia:2019bqz} or in Little Higgs 
models~\cite{Kilian:2004pp}). Further resonances in these models can show up in order to include dark 
matter~\cite{Alanne:2018xli}, and likewise in models of top-quark PC~\cite{Ferretti:2013kya}. There have been also 
studies on the reach at a future 100 TeV FCC-hh collider for technicolor $\rho$ resonances (from Drell-Yan via mixing or VBF) or scalar 
$\sigma$ resonances, showing that the mass reach extends to roughly 15-20 TeV~\cite{Franzosi:2016aoo,Thamm:2015zwa}. 

There is a plethora of different models following the par\-a\-digm of PC~\cite{Ferretti:2016upr}, which differ by the 
coset space of the global symmetries, restrictions on the number of hyper-color multiplets, implementations,  charges 
of Abelian and flavor symmetries, and the specific hypercolor fermion representations. For example: there must not be 
too many fermion multiplets because otherwise the hypercolor group would not condense. As mentioned above, flavor and 
EWPO data constrain the misalignment angle. These constraints then turn into theoretical predictions and limits for 
multi-boson processes like diboson or VBS topologies. Most of the VBS studies for composite Higgs models have been done 
in the EFT framework (\eg~\cite{Grober:2010yv}, parameterized in terms of $\xi \equiv v/f$ mapped to the SILH 
Lagrangian~\cite{Giudice:2007fh}). As shown in~\cite{Belyaev:2012bm}, scattering processes involving 3, 4, or more 
external EW bosons (including the 125 GeV scalar state) are very good to search for discrepancies between the SM and 
composite Higgs models. Double and even triple Higgs production in VBF has been studied for LHC and higher-energy 
proton colliders. Similar processes have been also investigated for high-energy lepton 
colliders~\cite{Contino:2013gna}. For a proper study of complete composite Higgs models, additional composite 
resonances have to be added to the simulation of the VBS processes. Lastly, photon-induced processes using large photon 
fluxes at high-luminosity lepton and hadron colliders, \eg\ from processes like $\gamma\gamma \to HH$, can also be used to search for deviations, or directly as a probe of new physics~\cite{Alva:2014gxa,Fuks:2019clu}.

Though composite and Little Higgs models in general lead to extensions of the SM scalar sector, the composite nature of new resonances make these models special cases. On the other hand, there are scalar extensions with fundamental particles which also have connections to multi-boson physics and VBS~\cite{RSantos}. Also here, there is a gigantic model space: 
In principle, there can be an arbitrary number of (additional) Higgs singlets, Higgs doublets, and Higgs triplets (or 
even higher weak isospin representations);  models can be embedded into supersymmetric setups; coefficients of the 
potential can be real or intrinsically complex; there can be additional discrete symmetries; and models differ on how 
additional Higgs multiplets decouple from the SM fermion sector. These additional states can cause mixing effects on 
the different scalar Higgs states and CP admixtures. In almost all these cases there are additional neutral scalars, 
and in all multi-doublet Higgs models there are charged Higgs bosons.

One of the largest constraint for non-minimal models (like 2HDM) is the $\rho$ parameter, which parameterizes  the radiative corrections to the ratio $M_W / (\cos\theta_W M_Z)$. There are also several constraints on models with such 
extended scalar sectors: tree-level unitarity for scattering processes of  SM as well as new particles, the boundedness 
of the scalar potential from below, the existence of an absolute minimum in the potential. Some of them are 
automatically fulfilled in 2HDM, but not in more general Higgs multiplet models. Usually, 2HDMs are classified according to the discrete symmetries obeyed by the fermion Yukawa couplings to the doublets~\cite{Barger:1989fj,Branco:2011iw}. The classification labels Type I, II, etc., are related to the precise set of discrete symmetries imposed on the couplings.
Among the most prominent searches for the 2HDM at the LHC is the search for 
pair production of heavy Higgs bosons with decays to lighter Higgs bosons and $b$ quarks. While the exclusion limits depend 
crucially on the mixing angles of the vacuum expectation values of the Higgs doublets ($\tan\beta$) and how far the model is in the decoupling limit 
(where the states of the two Higgs doublets are ordered as a completely SM-like $h(125)$ state as a custodial SU$(2)_c$ 
singlet and a rather degenerate SU$(2)_c$ triplet of heavy states $H, A, H^\pm$), these searches can be used for all variants of the 2HDM.
This is irrespective of the couplings to the fermion sector. Measurements of the Higgs couplings can be used 
to detect admixtures of other states in the $h(125)$, \eg\ in non-minimal 2HDM (N2HDM type II, as 
in~\cite{Muhlleitner:2016mzt}), particularly to the admixture of Higgs singlets. The actual sign of each $\kappa$ in 
the $\kappa$ framework for Higgs coupling measurements depends on the chosen range of Higgs mixing angles in 
2HDM~\cite{Ferreira:2014naa,Ferreira:2014dya}. In the 2HDM, there is a strong correlation between the Higgs coupling 
measurements and the possible discovery reach in VBS: perturbative unitarity of the 2HDM imposes a sum rule on the 
couplings of all the scalar (and pseudoscalar) states to two EW bosons. The stronger the $HWW$ and $HZZ$ are 
constrained by the Higgs coupling measurements, the more SM-like the unitarization of VBS due to the $h(125)$ is. 
Future lepton colliders like ILC and CLIC  allow for a very precise determination of Higgs couplings and have the 
highest discovery potential for these models in the near future~\cite{Azevedo:2018llq}.
Another interesting topic is the question of possible CP violation in the scalar sector. Conclusive measurements at the 
LHC are rather difficult, and this most likely also needs a future lepton collider. 

Next, we discuss Higgs triplets and the GM model. As there is a destructive interference between the 
hypercharge and the weak contributions to the $HVV$ couplings one needs a scalar with at least isospin $T_L=1$ to 
enhance the SM rate. Furthermore, additional states need to get vacuum expectation values and they need to mix with 
$h(125)$. These models have both interesting symmetry-breaking patterns, as they are able to generate neutrino masses, 
as well as phenomenology, which for example, can enhance the $h \to \gamma\gamma$ decay rate via doubly charged Higgs 
loops. A very popular implementation is the GM model, which features an isospin doublet $\Phi$ with 
$Y=1/2$, a complex triplet $\chi$ with $Y = 1$, and a real triplet $\xi$ with $Y= 0$. The interesting case is when the 
two triplets are aligned as then these models exhibit an accidental SU$(2)_c$ custodial symmetry and $\rho = 1$ at tree
level. The parameter space of this model has been studied in~\cite{Hartling:2014zca} where deviations in $\kappa_V$ and 
$\kappa_f$ have been scanned. 

At the workshop, the experimental search for singly and doubly charged Higgs bosons at 
the LHC was discussed~\cite{LBarak}. For light, charged scalars, there are searches for decays $H^+ \to c\bar{s}$ as well $H^+ \to W^+ a$, where 
$a$ is a light pseudo-scalar Higgs boson, and also $H^+ \to t^{(*)} \bar{b}$ (potentially off-shell top decay). These 
search channels are mostly used for heavier or heavy charged Higgs bosons, where now in addition there is the decay 
channel $H^+ \to W^+ h(125)$. Also VBF charged Higgs production with $H^+ \to W^+\gamma$ has been investigated~\cite{Sirunyan:2019zdq}. During 
early data-taking, pair production of doubly charged Higgs bosons $(H^{\pm\pm})$ via the Drell-Yan mechanism decaying 
to pairs of (light) same-sign leptons has been the dominant search signature~\cite{Aaboud:2018qcu}. Under the considered model assumptions, 
masses of doubly charged Higgs bosons below $m_{++}=700-800$ GeV have been excluded. More recently, single production
of $H^{\pm\pm}$ through same-sign $WW$ fusion has been considered. Due to a smaller signal rate, however, only 
$m_{++}\lesssim 250$ GeV has been excluded. Both searches can benefit from increased statistics. Complementary to this experimental effort,  there is still much 
room for improvement in MC modeling of singly and doubly charged Higgs production at hadron 
colliders~\cite{RRuiz,Fuks:2019clu}. Using state-of-the-art MC tools, modeling prescriptions and recommendations were 
presented covering: the Drell-Yan channel at NLO+PS and with jet vetoes; $\gamma\gamma$ fusion at LO+PS and with a
systematic assessment of photon PDF uncertainties; $gg$ fusion with NNNLL threshold resummation; and VBF at NLO+PS,
including generator-level cuts within the MC@NLO formalism~\cite{Fuks:2019clu}. A key link in this modeling was the development of the NLO UFO libraries {\texttt{TypeIISeesaw}}, which are publicly available from the {\texttt{FeynRules}} model database.

\subsection{Specific topics in precision of SM VBS processes}
\label{sec:sm_vbs_proc}

Though the main topic of the Lisbon COST ``VBScan'' workshop was the connection of VBS to BSM models, no deviation from the SM can be reliably established without precise knowledge on the SM predictions. Indeed, the COST ``VBScan'' network initiated, through a series of workshops, a comparative study on NLO and LO precision prediction at parton level and matched to parton showers for same-sign VBS~\cite{Ballestrero:2018anz}. More references to the dedicated papers can be found therein.

The difficulty is to restrict the full off-shell VBS process to the underlying VBS dynamics. This must not be done by selecting subsets of Feynman diagrams, as VBS and triboson production are together in an inseparable gauge invariance class. For lepton colliders this mixes the VBS process with two undetectable forward neutrinos with the corresponding $VVZ$, $Z\to\nu\bar{\nu}$ final state~\cite{Boos:1997gw,Boos:1999kj,Fleper:2016frz}. The equivalent for hadron colliders like the LHC is the corresponding hadronic decay of one of the final state EW bosons to two jets, interfering with the two forward jets of the VBS subprocess. To isolate the underlying dynamics, one has to find a cut flow to a fiducial phase space enriching the VBS contribution. For lepton colliders this fiducial phase space is defined in Ref.~\cite{Boos:1997gw}, while for hadron colliders it is given in Ref.~\cite{BuarqueFranzosi:2019boy}.

\section{Synergies between theory and experimental analyses}
\label{sec:synergy}

The workshop concluded with an open discussion on how to practically implement experimental analyses that target new and previously-neglected BSM models or that can be exploited in a global fit of EFT operators simultaneously with other final states. It was noted that new and innovative ideas take time to propagate into experimental analyses. For example: well-known experimental analyses such as VBF Higgs production with invisible decays~\cite{Sirunyan:2018owy,Aaboud:2018sfi}, or dark matter production via VBF, may already be relevant for pNGBs dark matter~\cite{Ruhdorfer:2019utl}, but these models have not been considered by the experimental collaborations. Several theorists expressed a general desire for a better way to communicate models of interest to the experimental community. It was also noted that making public additional material to more readily allow reinterpretation of data is also strongly appreciated~\cite{Abdallah:2020pec}.

For experimental analyses targeting EFTs, it was emphasized that a global and consistent fit for dimension-6 operators across many final states is highly sought after by the EFT community. While such approaches will take time to implement, a critical mass of the experimental community is seemingly open to the proposal, and efforts have begun to move forward with this approach. It was, however, noted that many questions of sensitivity to different operators across final states could already be answered with a thorough MC study. For example, it has been demonstrated that dimension-6 operators can play a role compared to dimension-8 in VBS signatures~\cite{Gomez-Ambrosio:2018pnl}, but the role of VBS measurements in a global fit is not yet extensively known. An extensive MC study could take place on much faster time scale than an exhaustive set of experimental analyses, and would provide valuable input to  experimental efforts.

Further discussion focused on new techniques to expand searches for new physics. Trigger "scouting," used successfully by the CMS Collaboration to extend searches for new physics at low mass~\cite{Sirunyan:2019wqq}, offers an exciting avenue to explore unconventional or experimental difficult signatures. With the huge data sets expected from the HL-LHC phase with upgraded  detectors, significant improvement in reconstruction algorithms is envisioned. As discussed in the dedicated talk~\cite{GStrong}, ML is expected to play a critical role in these advancements.

\section{Conclusions}
\label{sec:summ}

Vector boson scattering (VBS) and triple boson production measurements at the LHC until now have been more or less exclusively been in the realm of the SM and EW working groups of the experimental collaborations at the LHC. Deviations from the SM in these channels have been searched for in the rather generic framework of effective field theories (EFTs). The main purpose of this workshop was to bring together experts on model building and phenomenologists working on BSM scenarios relevant for VBS (and tribosons) with the experimental community working on these channels. The intention was to kick off and foster collaborations between the two different communities, to provide an overview to the experimental groups on the relevant model space and to present these specific rarest EW processes at the LHC as vehicles for BSM searches. 

There were many interesting discussions taking place between the talks, after the sessions, and also during the social by-programme of the workshop. 
These discussions covered many different topics, \eg\ the question on the importance of global fits for EFTs vs. limits on individual Wilson coefficients, the question of complete bases for EFT, and of course the proper definition for signal models for high-energy tails in this setup. For all different VBS channels there were constructive and continuous discussions on the available SM theory precision predictions for these processes and the corresponding theory uncertainties, particularly in the high-energy bins, and also on the availability, flexibility, and physics coverage of different theory tools.
There was also an extended discussion on the upcoming stronger inclusion of semi-leptonic and even fully hadronic measurements in the VBS channels, as those processes are the only ones, together with the rare $ZZ \to \ell\ell\ell\ell$, that allow to reconstruct the full invariant mass of the diboson system and hence the energy scale of EFT operators to be constrained. 

An extensive discussion was devoted to the coverage of the searches and the question whether there is a non-negligible model space that is not searched for in VBS (or in general by the experiments). This led to a longer discussion on triggers, on the question of definition of final states, and the coverage via fiducial phase space volumes, and in general the question on model-independence and searches without bias.

On the experimental front, studies of VBS processes in the SM are expanding rapidly. The understanding of $ZZ$, $WZ$, and $W^\pm W^\pm$ production from VBS have all  been advanced by exploiting the full Run~2 data set. Experimental techniques for reconstruction and selection, including machine learning, have also allowed channels with more challenging hadronic decays to become accessible. As the full Run~2 data set is analyzed in all channels, and with the luminosity of the LHC Run~3, VBS is an exciting experimental probe. 

The focus of experimental efforts in VBS have largely focused on the first observations and characterization of SM VBS production. In many cases, constraints on anomalous VBS production in the language of EFT have also been placed. However, dedicated searches for specific models that are most sensitive to VBS production are less common. The theoretical component of this workshop provided important context and considerations for the validity and applicability of EFT constraints, as well as many ideas for areas where VBS searches would be welcome. Improving the breadth of experimental searches to new, unexplored models can significantly expand the role of VBS studies in characterizing extensions of the SM.

Finally, we hope that this write-up serves as a collection of material for the status of VBS and triboson processes at the LHC together with an overview over a selection of BSM models for which these search channels might prove interesting or relevant. 

\section*{Acknowledgments/Impressum}

This workshop was organized by the COST Action CA16108, which funded the overwhelming majority of this workshop.
We also acknowledge the support of the Funda\c{c}\~ao para a Ci\^encia e a Tecnologia, Portugal.\\

This work is licensed under the Creative Commons Attribution 4.0 International License. To view a copy of this license, visit CCBY 4.0 or send a letter to Creative Commons, PO Box 1866, Mountain View, CA 94042, USA. \qquad\qquad\qquad
\includegraphics[width=.06\textwidth]{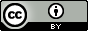}

Preprint report numbers:
DESY-PROC-2020-002, \;
ISBN 978-3-945931-33-2, \;
ISSN 1435-8077\; \\
CP3-20-17, VBSCAN-PUB-04-20


\bibliography{biblio}

\end{document}